# Experimental demonstrations for randomness-based macroscopic Franson-type nonlocal correlation using coherently coupled photons

S. Kim and Byoung S. Ham[*]
School of Electrical Engineering and Computer Science, Gwangju Institute of Science and Technology
123 Chumdangwagi-ro, Buk-gu, Gwangju 61005, S. Korea
(Submitted on July 1, 2021; [*]bham@gist.ac.kr)

**Abstract**
Franson-type nonlocal quantum correlation based on the particle nature of quantum mechanics has been intensively studied for both fundamental physics and potential applications of quantum key distribution between remotely separated parties over the last several decades. Recently, a coherence theory of deterministic quantum features has been applied for Franson-type nonlocal correlation [arXiv:2102.06463] to understand its quantumness in a purely classical manner, where the resulting features are deterministic and macroscopic. Here, nearly sub-Poisson distributed coherent photon pairs obtained from an attenuated laser are used for the experimental demonstrations of the coherence Franson-type nonlocal correlation. As an essential requirement of quantum mechanics, quantum superposition is macroscopically provided using polarization basis-randomness via a half-wave plate, satisfying fairness compared with the original scheme based on phase bases. The observed coherence quantum feature of the modified Franson correlation successfully demonstrates the proposed wave nature of quantum mechanics, where the unveiled nonlocal correlation is relied on a definite phase relation between the paired coherent photons.

**Introduction**

Over the last several decades, the Bell theorem [1-5] has been a major guideline and a testing tool for nonlocal correlations in quantum mechanics [6,7]. Back in 1935, Einstein, Podolsky, and Rosen raised a fundamental question on quantum mechanics, the so-call EPR paradox, regarding nonlocal quantum correlation between paired particles remotely separated [8]. In quantum mechanics, a particle (photon, spin, or atom) can be described by using Schrodinger's wave equation for its wave nature or an energy (Fock) state for its particle nature. According to the Copenhagen interpretation of quantum mechanics, a photon can be viewed as a particle or a wave in a mutually exclusive manner, resulting in the wave-particle duality [9]. This is the core concept of the complementarity theory [10]. Especially for photons, however, developments of quantum mechanics over the last millennium are mostly for its particle nature, where the wave nature has been nearly excluded. A new interpretation on the wave nature-based quantum correlation has been recently proposed to view quantum mechanics deterministically and macroscopically in several ways [11-13]. Such an idea has also been experimentally demonstrated [14,15]. In the coherence approach, the phase information between interacting particles is the bedrock for the deterministic quantum information processing. Here, we report the first experimental demonstration of the wave nature-based Franson-type nonlocal correlation using coherent photon pairs from an attenuated laser [13], where the coherent photon pairs are nearly perfectly sub-Poisson distributed.

To experimentally demonstrate quantum entanglement, two major test tools have prevailed. One is a Hong-the Ou-Mandel (HOM) correlation test [16-18], and the other is the Bell inequality violation test [1-5]. Wigner's distribution has also been a major test tool with a negative feature of photon distribution regarding observer matters [19-23]. However, it is also true that the positive nature of Wigner's distribution does not violate quantum nature of entanglement [22]. In other words, the nonnegative feature of Wigner's distribution also satisfies quantum property of entanglement. The HOM method relies on an interferometric system of a beam splitter (BS), where an impinging photon pair results in photon bunching or anticorrelation on a BS due to destructive quantum interference [16]. According to a recently proposed new interpretation for the HOM correlation using the wave nature of a photon, such a quantum feature can be deterministic according to the phase relation between the paired photons [11,14]. In this sense, spontaneous parametric down conversion (SPDC)-generated entangled photon pairs should also satisfy such a phase relation with a π/2 phase shift [23]. More recently, such a π/2 phase relation between entangled photons has been experimentally demonstrated using attenuated coherent photons, satisfying nearly sub-Poisson distribution [14]. In the wave nature of a photon, the destructive quantum interference of anticorrelation is self-evident due to an additional $\pi/2$ phase shift induced by a BS [24].

Like the HOM-type quantum correlation, the Franson-type non-local correlation [25-28] has also been newly interpreted using the wave nature of a photon [13]. The Franson correlation is an interferometric version of Bell inequality using noninterfering MZIs [26-29]. The observed two-slit interference-like fringe is an inherent property of the Franson correlation. Compared with the original asymmetric Mach Zehnder interferometers (MZIs) based on the phase basis [26], the symmetric MZIs in Fig. 1 can also satisfy such a Franson-type requirement by using polarization bases [29]. According to the Frensel-Arago law, orthogonal bases do not interfere on a BS, resulting in no interference fringe [30]. For the quantum correlation between the non-interfering MZIs, path randomness can be introduced to both MZIs by using half-wave plates, where the randomness is a fundamental requirement for quantum superposition, resulting in indistinguishability [31,32].

**Results**

*A modified Franson scheme*

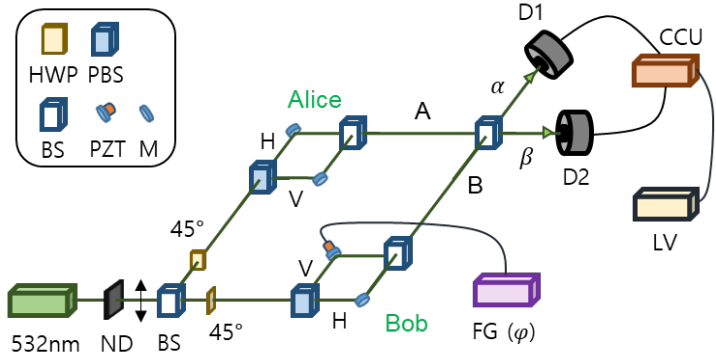

**Fig. 1| Schematic of Franson-type nonlocal correlation based on polarized coherent photons.** D: a single photon detector. CCU: coincidence counting unit. LV: Lab View software. H (V): horizontal (vertical) polarization. BS: beam splitter, M: mirror, PZT: Piezo-electric transducer, HWP: half-wave plate, PBS: polarizing BS.

Figure 1 shows a coherence version of the Franson-type nonlocal correlation based on the wave nature of coherent photons [13], where the coherent photons are provided to be nearly sub-Poisson distributed from an attenuated continuous wave (cw) laser at 532 nm (Coherent, Verdi-V10). The coherence length (time) of the cw laser is calculated to be ~60 m (~0.2 μs) from a given ~5 MHz spectral linewidth. Unlike the original Franson scheme of asymmetric MZIs based on phase bases of paired photons, Fig. 1 shows symmetric MZIs composed of polarizing BS (PBS) and nonpolarizing BS for orthogonal polarization bases, H (horizontal) and V (vertical), resulting in the same function of noninterferometric Franson scheme. For the relative phase control between two MZIs, we set a reference for Alice's phase $\psi$, whereas Bob's phase ($\varphi$) is deterministically controlled by a piezoelectric transducer (PZT; Thorlabs, POLARIS-K2S3P) via a PZT controller (Thorlabs, MDT693B). The PZT controller is operated by a function generator (Tecktronics AFG3102). For this, the PZT-induced phase $\varphi$ is repeatedly scanned for a fixed few micron PZT scan range. The 45° half-wave plate (HWP) in Fig. 1 rotates the vertically polarized original photons into a diagonal direction for random polarization basis for the following MZI, where the rotation angle of HWP is actually 22.5° from the major axis of the birefringent crystal. Thus, the photons entering PBS-BS MZIs in both parties satisfy the random (indistinguishable) polarization basis required for quantum superposition. This quantum superposition in each MZI is the origin of the Franson interference fringe between MZIs.

In the original scheme of the Franson correlation, each asymmetric BS-BS MZI is composed of short and long paths, whose path-length difference is much longer than the coherence length of the interacting photons [26] (see Section A of the Supplementary Information). On the contrary, the symmetric MZI configuration in Fig. 1 works for orthogonally polarized photons, resulting in the same function of the non-interferometric interferometer. The MZI path randomness for quantum superposition of a nonpolarized single photon is satisfied by a beam splitter (BS) in the original scheme. On the contrary, the PBS in Fig. 1 plays the same role of quantum superposition for the orthogonally polarized photons. Phase coherence between Alice's and Bob's MZIs plays a major role for the interference fringe of the Franson correlation in coincidence detection measurements. Such phase coherence is also provided by photon pairs generated from the spontaneous parametric down conversion process (SPDC), whose phase difference between the paired photons is fixed at $\pi/2$ [11,33]. Coherent photons from an attenuated laser automatically satisfy the in-phase relation between photons generated by cavity optics. These four distinctive comparisons between the original and modified Franson schemes are summarized in Table 1.

Table 1. Comparison between the original scheme and Fig. 1.

|  | Original Franson setup [26] | Fig. 1 |
|---|---|---|
| Non-interferometric scheme | Asymmetric MZI | Symmetric MZI |
| Basis | Phase $\{0, \pi\}$ | Polarization $\{H, V\}$ |
| MZI path Randomness | Nonpolarizing beam splitter | Polarizing beam splitter |
| Basis coherence | SPDC | Cavity optics |

The attenuated photon characteristics from the 532 nm cw laser in Fig. 1 confirm the nearly sub-Poisson distribution with a mean photon number of $\langle n \rangle \sim 0.04$ (see Section B of the Supplementary Information). In Fig. 1, coherently paired photons are post-selected via coincidence detection measurements (see Fig. 2), where the doubly-bunched photon generation rate is ~1% of the single photon generation rate. The three bunched photon generation rates are also confirmed to be ~1% of the doubly-bunched rate, resulting in an overall error of ~1% in the coincidence measurements. In our experiments, the single photon detectors (D1 and D2) do not resolve photon numbers.

*Theoretical analysis*

Before discussing the experimental results of Fig. 1, we first need to set a reference based on the original scheme of the Franson correlation to analyze the new data based on the polarization bases. From the original Franson scheme of asymmetric MZIs [26], the following is obtained (see Fig. S1(a) of the Supplementary Information):

$$|\Psi\rangle_A = (|S\rangle_A + ie^{i\varphi}|L\rangle_A)/\sqrt{2}, \qquad (1)$$
$$|\Psi\rangle_B = ie^{i\theta}(|S\rangle_B + ie^{i\psi}|L\rangle_B)/\sqrt{2}, \qquad (2)$$
$$\langle R_{AB} \rangle = \frac{I_0^2}{2}\langle 1 + \cos(\varphi - \psi)\rangle, \qquad (3)$$

where $I_0 = E_0 E_0^*$, and the subscripts A and B represent the photons to Alice and Bob, respectively. $\langle R_{AB} \rangle$ in equation (3) is the coincidence measurements between the A and B output paths from both MZIs. Due to the noninterfering MZIs, the product term $\langle S|L\rangle_{AB}$ is cancelled, where $\langle R_{AB} \rangle = \langle |\langle \Psi|\Psi\rangle_{AB}|^2\rangle$. $\Psi$ and $\varphi$ are the relative phase differences of each MZI between the short and long paths, respectively. Due to random detuning of the signal and idler photons in $\chi^{(2)}$-based SPDC, the mean intensities of equations (1) and (2) are uniform regardless of $\varphi$ and $\psi$: $\langle\langle\Psi|_A|\Psi\rangle_B\rangle = \langle\langle\Psi|_B|\Psi\rangle_A\rangle = 1$ $(I_0)$. As expected in equation (3), however, coincidence detection results in an interference fringe as a function of the phase difference between both MZIs. This noninterfering interference fringe is an intrinsic property of the Franson-type nonlocal correlation [26].

Except for the final BS (see the dotted circle) for A and B interference in Fig. 1, exactly the same configuration is provided as the original Franson scheme, where S and L are replaced by H and V, respectively [13]. Due to the noninterfering scheme, $\langle H|V \rangle_{AB} = 0$ by the Fresnel-Arago law [26,30]. The phase information between two paired photons in the Franson-type nonlocal correlation can also be tested for a typical interference scheme of the Hong-Ou-Mandel type correlation by inserting the final BS for the outputs A and B in Fig. 1 [13]:

$$I_\alpha = \frac{I_0}{2}[2 - \cos(\theta) - \cos(\varphi - \psi - \theta)], \qquad (4)$$
$$I_\beta = \frac{I_0}{2}[2 + \cos(\theta) + \cos(\varphi - \psi - \theta)]. \qquad (5)$$

Equations (4) and (5) offer richer information of the quantum feature than equation (3), where the global phase $\theta$ between A and B affects both output fields (see Fig. 2). The global phase, $\theta$, in equation (3) has no effect on $\langle R_{AB} \rangle$. The phase information of the paired photons may not be detected by our coincidence counting unit (CCU; Altera DE2), where the CCU speed is much slower than that of single photon counting modules (SPCMs: Excelitas AQRH-SPCM-15). This is the practical reason for using the interference scheme by inserting the last BS to interfere $\alpha$ and $\beta$. The results are compared with those of the original Franson correlation for richer physics of quantum mechanics below.

*Experimental analysis*

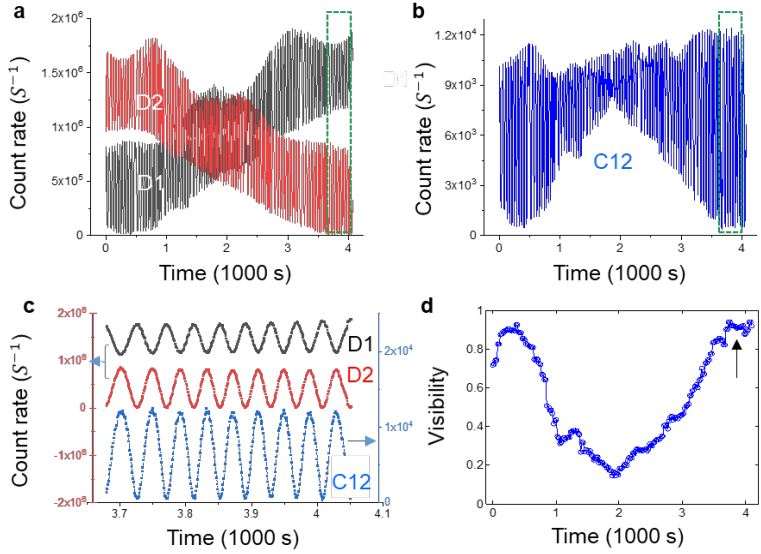

**Fig. 2| Experimental results for Fig. 1. a** Individual output counting rate for $\alpha$ (D1) and $\beta$ (D2). **b** Coincidence counting rate C12 for the output channels D1 and D2. **c** Expansion of (**a**) and (**b**) for the green dotted boxes: color matched. **d** Visibility between for $\alpha$ (D1) and $\beta$ (D2). The fast oscillation is due to the PZT-controlled phase $\varphi$, while the slow modulation is due to the global phase $\theta$ between A and B in Fig. 1.

Figure 2 shows experimental data of the coherent photon-based Franson-type correlation for Fig. 1. For the global phase $\theta$-effect in equations (4) and (5), we conduct experiments for ~4,000 s (~66 minutes), where the air turbulence-caused phase fluctuation is affected naturally under normal lab conditions, resulting in a slow modulation in each output photon measurement, as shown in Fig. 2a. The PZT-scan-caused $\varphi$-dependent fast modulation is the key result of the Franson correlation in both outputs (see Fig. 2c), whose fringe period is the wavelength of the 532 nm laser. The PZT is scanned repeatedly at a 1 mHz speed via the function generator, resulting in total four full repeated scans (see Methods and Section C of the Supplementary Information).

In Fig. 2, the MZI setup is enclosed by a cotton box to reduce air turbulence, where the size of each small PBS-BS MZI is about a quarter of the large BS-BS MZI in Fig. 1. Each fast modulation fringe caused by the interference between two small PBS-BS MZIs is also affected by the air turbulence globally, resulting in a local variation of the overall fringe (see Fig. 2a) as well as visibility (see Figs. 2b and 2d). Such a global phase-dependent slow modulation has never been reported in the original Franson-type nonlocal correlation so far [25-28]. This is the first richness of the present scheme based on the wave nature of photons for the quantum correlation.

Figure 2b shows the coincidence detection signal (C12) between two outputs, $\alpha$ and $\beta$, where the coincidence detection has the same pattern as the product of D1 and D2 (see Fig. 3c). As mentioned above, the ratio of doubly bunched photons (Fig. 2b) to single photons (Fig. 2a) is ~1%. Unlike the SPDC-generated wide bandwidth at ≫ GHz, the present coherent laser has an extremely narrow bandwidth of ~5 MHz, so that the ~350 ps resolving time of SPCMs does not have a negative effect on the measurement results.

Figure 2c is an extension of the green dotted boxes at t~3900 s in Figs. 2a and 2b. The D1 and D2 in Fig. 2c denote equations (4) and (5) for $\theta = 0$ (or $2n\pi$), respectively (refer to the left axis). The coincidence measurements (C12) refer to the right axis, where C12 has the same fringe period as in equation (3) as well as the original Franson correlation [26]. The signal C12 in Fig. 2c also demonstrates nearly perfect visibility for the present scheme of Franson correlation, where the global phase is set at $\theta = \pm 2n\pi$ (see the arrow in Fig. 2d). For the visibility calculations in Fig. 2d, a homemade Matlab program was used to evaluate the data of Figs. 2a and 2b, where the visibility decreases as the global phase is away from $2n\pi$. For $\frac{\pi}{4} < \theta < \frac{3\pi}{4}$, it shows a classical feature (analyzed in Fig. 3).

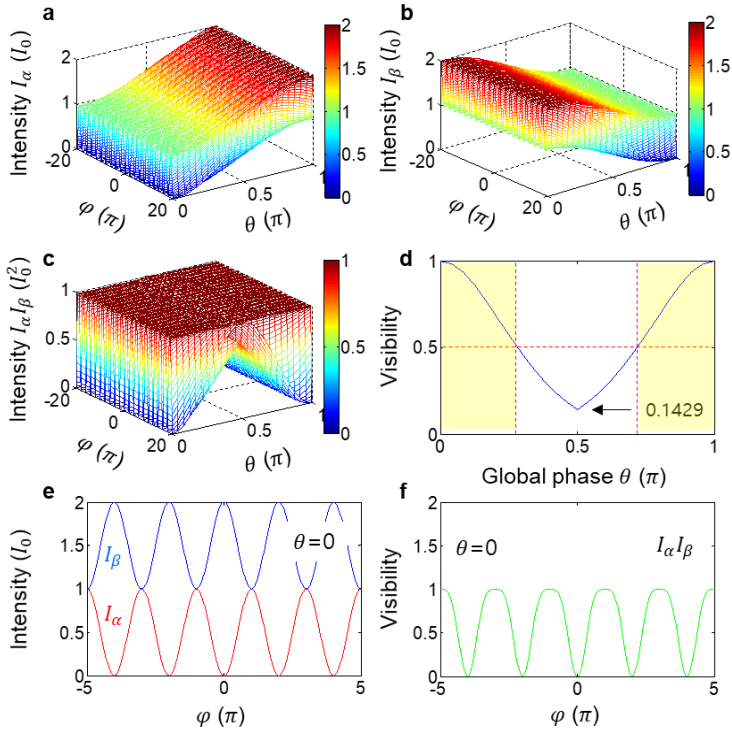

**Fig. 3| Numerical calculations for Fig. 2. a,b,c** Output intensities $I_\alpha$, $I_\beta$, and $I_\alpha I_\beta$. **d** Visibility between $I_\alpha$ and $I_\beta$. The yellowish areas are for the quantum feature. **e** Details for (**a**) and (**b**). **f** Details for (**d**).

Figure 3 shows the numerical calculations for equations (4) and (5), which correspond to Fig. 2. For these calculations, Alice's MZI is fixed at $\psi = 0$, while Bob's MZI is φ-controlled by PZT. To comply with natural air turbulence as shown in Fig. 2a, we set slow θ scanning with respect to the fast φ-scanning. Figures 3a and 3b show the corresponding output intensities of α and β, respectively, resulting in nearly the same feature as in Fig. 2a. Figure 3c shows the product of Figs. 3a and 3b as functions of both θ and φ, resulting in the same feature as in Fig. 2b for the coincidence detection. Thus, the coincidence detection between coherently paired photons observed in Figs. 2b is understood as a result of entangled photon interactions on a BS. Because photons never interfere with others as claimed by Dirac [34], Figs. 2b and 3c are due to self-interference and fully support the phase relation between them. In other words, the interpretation of the quantum features of the Franson-type correlation is now clearly understood for the wave nature of a photon, resulting in the same feature as conventional correlations based on the particle nature.

Figure 3d shows the visibility for both output fields of $I_\alpha$ and $I_\beta$ obtained in Figs. 3a and 3b for all $\varphi$s as a function of $\theta$. The calculated visibility in Fig. 3d is similar to Fig. 2d under the air turbulence-caused errors, where the nonclassical feature of the Franson correlation is maximized (minimized) at $\theta = \pm n\pi$ $\left(\theta = \pm \frac{(2n+1)\pi}{2}\right)$. This is obvious in an interferometric system because at every $\pi$, $I_\alpha$ and $I_\beta$ are swapped. At $\theta = \pm \pi/2$, the paired photons A and B lose entanglement, resulting in no photon bunching as shown in the Hong-Ou-Mandel analysis (see Section D of the Supplementary Information) [33]. Even for a $\theta$-phase locked condition, the visibility is reached at V=0.14 if $\theta = \pm \pi/2$. This fact has never been discussed or observed in conventional Franson-type nonlocal correlation. Figures 3e and 3f exactly match Fig. 2c. Thus, the experimental results of Fig. 2 for the coherence version of the Franson-type nonlocal correlation strongly support the wave nature-based quantum mechanics previously reported [13]. The quantum feature of the Bell inequality violation is not mysterious or spooky, but deterministic according to the phase relation between the paired photons.

**Discussion**

Based on the Born rule for multi-particle-multi-slit interference, a photon entering an MZI is described only by two-slit quantum superposition, resulting in self-interference [35]. Thus, for a pair of photons, a four slit system is minimally required to fully describe the quantum interference phenomenon [36,37]. For the modified Franson scheme in Fig. 1, each photon enters each noninterfering MZI, resulting in self-interference in each output. Self-interference is a definite feature of the wave nature of a photon. According to the intensive researches conducted over the last decade, it is well known that there is no difference between a particle and a wave for the interference results [36]. This fact was already mentioned by Dirac in 1950s that photons never interfere with others [38]. In that sense, the Franson correlation between two output photons from independent MZIs is originated in the phase relation between the originally paired photons. Such phase relation is definitely detected by the last BS in Fig. 1 regardless of the distance between the paired photons. Even without the BS for the original Franson scheme, the coincidence detection between the output photons must be coherent, otherwise the controlled phase information in each MZI cannot be read out for an interference fringe. The major difference between the original and the modified Franson schemes is whether to read the global phase ($\theta$) effects. This $\theta$ effect contributes to enrich our understanding on quantum correlation that has been vailed by coincidence measurements. In that sense, a coincidence detection module needs to be tested for the phase information transfer from an optical signal to an electrical signal. Such phase coherence transfer has already been demonstrated in nonlinear quantum optics using acousto- and/or electro-optic modulators driven by a synchronized microwave field to an optical field [39].

**Conclusion**

We experimentally demonstrated the coherence version of Franson-type nonlocal correlation proposed in ref. 13 in an interferometric scheme using nearly sub-Poisson distributed coherently coupled photon pairs obtained from an attenuated cw laser at 532 nm. Unlike conventional Franson-type nonlocal correlation using SPDC-generated entangled photon pairs in an asymmetric MZIs, the present demonstration was based on polarization basis randomness for symmetric MZIs. Like conventional Franson-type nonlocal correlation, the observed data of coincidence detection also showed the same interference feature with even nearly perfect visibility. With the wave nature-based interpretation of quantum mechanics, thus, a deeper and richer understanding on the physical properties of the nonlocal correlation was achieved with unprecedented details such as for a global phase and phase-controlled superposition correlation. The theory and experimental data coincided nearly perfectly, demonstrating a prospective feature of the wave nature-based quantum correlation. Thus, the present experimental demonstration strongly supports the new interpretation of the wave-nature-based quantum correlation and opens the door to a deterministic quantum information science controlling the relative phase basis of noninterfering photon pairs.

**Methods**

A nearly coherent single photon stream was obtained from an attenuated 532 nm laser (Verdi V10) using ND filters with OD 13, where the original laser power was 10 mW. Each detected single photons by SPDMs were transferred to the DE2 board, whose main processor is FPGA. Not only single photon counts for each channel, but also coincidence counts between channels were recorded simultaneously. To display and to record raw data of the counted photons automatically, a Labview program was used. Regarding polarization basis randomness for photons, the half wave plate was rotated for 22.5 degrees from the fast axis of the wave plate. Regarding PZT scan controls, a three-axis-controlled mirror mount (Thorlabs, KC1-PZ) was used. Although this method should affect overlapping between two input photons on a BS, actual misalignment is negligibly small not to affect overall correlation. A triangle waveform from the function generator (Tektronix, AFG3102) at 1 mHz scan speed was set for a sweep mode to induce a repetitive linear variation of PZT voltage between 0 V and 100 V, resulting in a few micron path-length change. Due to non-perfect fringe match with respect to the voltage outputs, there was a seam in the fringe pattern at each turning point toward backward scan (see Fig. S4 of the Supplementary Information). As a result, the actual counted data for each scanning of the PZT was 455, otherwise 500. Each data point in Fig. 2 is for 1 s acquisition.

*Numerical simulations*: For the numerical calculations in Fig. 3, the analytic solutions of equations (4) and (5) were directly used for a homemade Matlab program. For the visibility V, a general definition was applied at each data point: $V = \frac{I_{max} - I_{min}}{I_{max} + I_{min}}$.

In a dark room condition enclosed by a black cartoon box, both single photon counting modules (Excelitas AQRH-SPCM-15) are tested for the dark count rate of $27 \pm 5$ (counts/s) without input photons, which is satisfied with the manufacture specification of 50 (count/s).

The SPCM-generated electrical pulses are sent to the coincidence counting module (CCU: DE2 FPGA, Altera). The pulse duration of each electrical signal from SPCM for single photon detection is ~10 ns. Both single and coincidence counting numbers are counted in parallel by DE2 for 1 s acquisition time for each data point, otherwise specified. The measured data is transferred to Labview via RS232 cable for a coincidence_rs232(4_5).vi application program in real time.

**Data availability**

The data presented in the figures of this Article are available from the corresponding author upon reasonable request.

**Code availability**

All custom code used to support claims and analysis presented in this Article is available available from the corresponding author upon reasonable request.

**Acknowledgments:** BSH acknowledges that this work was supported by GIST via GRI 2021.

**Additional Information**

Supplementary information is available in the online version of the paper. Reprints and permissions information is available online at www.nature.com/reprints.

**Competing interests**

The authors declare no competing interests.

**Author contribution**

B.S.H. conceived the idea and experimental methods, analyzed experimental data with numerical simulations, and wrote the manuscript. SK conducted experiment. Correspondence and request of materials should be addressed to BSH (email: bham@gist.ac.kr)